\documentclass[prd,twocolumn, nofootinbib]{revtex4}
\usepackage{epsfig}
\usepackage{subfigure}

\newcommand{\ltsim}{\mathrel{\raise.3ex\hbox{$<$\kern-.75em\lower1ex\hbox{$\sim$}}}}
\newcommand{\gtsim}{\mathrel{\raise.3ex\hbox{$>$\kern-.75em\lower1ex\hbox{$\sim$}}}}
\newcommand{\be}{\begin{equation}}
\newcommand{\ee}{\end{equation}}
\newcommand{\bea}{\begin{eqnarray}}
\newcommand{\eea}{\end{eqnarray}}

\begin{document}

\title{The peculiar velocity field: constraining the tilt of the Universe}
\author{Yin-Zhe Ma}
\email{yzm20@cam.ac.uk}
\affiliation{Kavli Institute for Cosmology and Institute of Astronomy, University of
Cambridge, Madingley Road, Cambridge, CB3 0HA, UK.}
\author{Christopher Gordon}
\affiliation{Oxford Astrophysics,
Physics, DWB, Keble Road, Oxford, OX1 3RH, UK}
\author{Hume A. Feldman}
 \affiliation{Department of Physics and
Astronomy, University of Kansas, Lawrence, KS 66045, USA}

\begin{abstract}

A  large bulk flow, which is in tension with the  Lambda Cold Dark
Matter ($\Lambda$CDM) cosmological model, has been observed.
 In this paper, we provide a physically \textit{plausible}
 explanation of this bulk flow, based on the assumption that some fraction of
the observed dipole in the cosmic microwave background is due to an
intrinsic fluctuation, so that the subtraction of the observed dipole leads to
 a mismatch between the cosmic microwave background (CMB) defined
 rest frame and  the matter rest
 frame. We investigate a model that takes into account the relative
velocity (hereafter the tilted velocity) between the two frames,
and develop a Bayesian statistic to explore the likelihood of this
tilted velocity.

By studying various independent peculiar velocity catalogs, we find that: (1) the magnitude of the
tilted velocity $u$ is around 400 km/s, and its direction is close
to what is found from previous bulk flow analyses; for most catalogs analysed, $u=0$ is excluded at
 about the $2.5 \sigma$ level;
(2) constraints on the magnitude of the tilted velocity can
result in constraints on the duration of inflation, due to the
fact that inflation can neither be too long (no dipole effect) nor
too short (very large dipole effect); (3) Under the assumption of a
super-horizon isocurvature fluctuation, the constraints on the
tilted velocity require that inflation lasts at least 6 e-folds
longer (at the 95\% confidence interval) than that required to
solve the horizon problem. This opens a new window for testing
inflation and models of the early Universe from observations of large scale
structure.

\end{abstract}

\maketitle



\section{Introduction}The bulk flow, i.e. the streaming
motion of galaxies or clusters, is a sensitive probe of
the density fluctuation on very large scales. Recently there have been
several observations of a large amplitude of the bulk flow on
hundred Mpc scales, which are in conflict
with the predictions of the $\Lambda$CDM model \cite%
{Watkins08,Feldman09}.
The bulk flow is measured with respect a frame in which the CMB temperature dipole vanishes.
We define this to the {\em CMB rest frame}. It is usually assumed that the CMB
rest frame coincides with the {\em matter rest frame} which we define to
be the frame in which the velocities of matter in our horizon volume are isotropic.

 It is possible that the two
reference frames actually do not coincide with each other,
resulting in a ``tilted universe'' \cite{Turner91,Langlois96,
Kashlinsky08}. If the inflationary epoch lasted just a little more
than the 60 or so e-folds needed to solve the horizon problem, the
observable ``remnants" of the pre-inflationary Universe may still
exist on very large scales of the CMB. As a result, there could be
some fraction of the CMB dipole due to the intrinsic fluctuations
rather than observer's kinetic motion. Therefore, when the
observed dipole is subtracted from the galaxy peculiar velocity
data, the subtraction induces a mismatch between the CMB rest
frame and matter rest frame.
We define the {\em intrinsic} CMB dipole to be the CMB dipole measured
in the matter rest frame.
The recent finding of such a mismatch
between the directions of the observed CMB dipole and
reconstructed velocity dipole \cite{Erdogdu06,Lavaux08} suggested
the possible existence of the intrinsic CMB dipole on the sky.

Pre-inflationary fluctuations in a scalar field may produce an
intrinsic dipole anisotropy. In this scenario, the observed CMB
dipole would be a sum of  the dipole from our motion in
the matter rest frame and the intrinsic CMB dipole caused by a
large scale perturbation.
This intrinsic dipole can be produced by a large scale
isocurvature perturbation \footnote{In our context,
an isocurvature perturbation is distinguished from an adiabatic perturbation
 in that the ratios of the number of photons to baryons and
 cold dark matter particles is not spatially invariant.},
 but not a large scale adiabatic
perturbation \cite{Erickcek08}. Note that our local motion
relative to the matter rest frame is caused by  small scale
inhomogeneity (up to about the 100 Mpc scale) and will be
negligibly affected by the very large scale ($\gg 10$ Gpc)
perturbation that would cause a tilt effect.

 In this paper, we estimate the tilted
velocity ($\mathbf{u}$) between the two rest frames using galaxy peculiar velocity data. This
opens a new window on testing early-Universe models from
observations of large scale structure.

\section{Likelihood and mock catalogs} In order to use the
galaxy peculiar velocity catalogues (see the following
descriptions of the data), we need to model the velocity of
galaxies in different rest frames. For each galaxy velocity survey,
we first subtract off our local motion with respect to the CMB as
estimated from the CMB dipole.  However, if there is a
non-negligible intrinsic CMB dipole, the CMB defined rest frame
will not correspond to the matter  rest frame and thus there will be
a residual dipole in the galaxy peculiar velocity survey. To test
this we estimate the line-of-sight velocity $S_{n}$ of each galaxy
{\it n} in the CMB rest frame with measurement noise $\sigma
_{n}$. Suppose the CMB rest frame has a tilt velocity $\mathbf{u}$
with respect to the matter rest frame, then the line-of-sight
velocity of each galaxy with respect to the matter rest frame
becomes $p_{n}(\mathbf{u})=S_{n}-\hat{r}_{n,i}u_{i}$, where
$\hat{r}_{n,i}u_{i}$ is the projected component of the 3-D
Cartesian coordinate $\mathbf{u}$ onto the line-of-sight direction
of galaxy {\it n}. After subtracting out the \textquotedblleft
tilted velocity\textquotedblright , we model the galaxy
line-of-sight velocity with respect to the matter rest frame
as $p_{n}=v_{n}+\delta _{n}$, where $%
v_{n}$ is the galaxy line-of-sight velocity in the matter rest
frame, and $\delta _{n}$ is a superimposed Gaussian random motion
with variance $\sigma _{n}^{2}+\sigma _{\ast }^{2}$, where $\sigma
_{\ast }$ accounts for the 1-D small scale non-linear velocity.
It can also compensate for an incorrect estimation
measurement noise $\sigma_{n}$ \cite{Watkins08,Feldman09}.
Therefore, the covariance matrix of $p_{n}(\mathbf{u})$ becomes
\begin{eqnarray}
G_{mn} &=&\left\langle v_{m}v_{n}\right\rangle +\delta _{mn}(\sigma
_{n}^{2}+\sigma _{\ast }^{2})  \nonumber \\
&=&\left\langle \left( \mathbf{\hat{r}}_{m}\mathbf{\cdot v(r}_{m}\mathbf{)}%
\right) \left( \mathbf{\hat{r}}_{n}\mathbf{\cdot v(r}_{n}\mathbf{)}\right)
\right\rangle +\ \delta _{mn}(\sigma _{n}^{2}+\sigma _{\ast }^{2}),
\label{Gmn}
\end{eqnarray}%
in which the cosmic variance term is \cite{Gorski88}
\begin{eqnarray}
\left\langle \left( \mathbf{\hat{r}}_{m}\mathbf{\cdot v(r}_{m}\mathbf{)}%
\right) \left( \mathbf{\hat{r}}_{n}\mathbf{\cdot
v(r}_{n}\mathbf{)}\right) \right\rangle \nonumber \\ =
\frac{\Omega _{\text{m}}^{1.1}H_{0}^{2}}{2\pi ^{2}}\int dk P(k)
f_{mn}, \label{vel_noi1}
\end{eqnarray}%
where
\begin{equation}
f_{mn}(k)=\int \frac{d^{2}\hat{k}}{4\pi }\left( \mathbf{\hat{r}}_{m}\mathbf{%
\cdot \hat{k}}\right) \left( \mathbf{\hat{r}}_{n}\mathbf{\cdot \hat{k}}%
\right) \times \exp \left( ik\mathbf{\hat{k}\cdot }\left( \mathbf{r}_{m}-%
\mathbf{r}_{n}\right) \right), \label{fmn}
\end{equation}
which can be calculated analytically (Appendix \ref{windowfmn}).

Therefore, the likelihood of the tilted vector and the small scale velocity
dispersion $\sigma _{\ast }$ can be written as%
\begin{equation}
L(\mathbf{u},\sigma_{\ast})=\frac{1}{(\det G_{mn})^{\frac{1}{2}}}
 \exp \left(
-\frac{1}{2}p_{m}(\mathbf{u})G_{mn}^{-1}p_{n}(\mathbf{u}) \right),
\label{likelihood}
\end{equation}
where we fix the cosmological parameters at the WMAP 7-year
best-fit values ($\Omega_{\text{b}}=0.0449$,
$\Omega_{\text{c}}=0.222$, $h=0.71$, and $\sigma_{8}=0.801$
\cite{Komatsu10}).

We parameterize the velocity as
$\mathbf{u}=\{u,\cos(\theta),\phi\}$ where in Galactic coordinates
$\phi=l$ and $\theta=\pi/2-b$. In order for our marginalized prior
on $u$ to be uniform we set
Prior$(\mathbf{u},\sigma_{\ast})\propto 1/u^2$. Then the posterior
distribution of the parameters ($\mathbf{u},\sigma _{\ast }$)
given the data ($\mathrm{D}$) is $ \textrm{Pr}(\mathbf{u},\sigma
_{\ast }|\mathrm{D})\propto \textrm{Prior}(\mathbf{u},\sigma
_{\ast })L(\mathbf{u},\sigma _{\ast })\,. $
 Before we perform the likelihood analysis for
the real data, we have tested this likelihood from mock catalogs.
We input a set of fiducial values of
($\sigma_{\ast}$,$\mathbf{u}$) and do 300 simulations of the data.
The average likelihood of these simulations for each parameter
exactly peaks at the input values, with the appropriate width
determined by cosmic variance, instrumental noise and small scale
velocity dispersion.

\begin{figure}[tbp]
\begin{center}
\includegraphics[width=8cm]{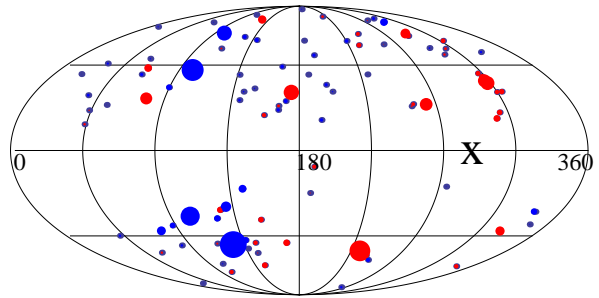}
\end{center}
\caption{The SN data plotted in Galactic
coordinates.. The red points are moving away from us and the blue
ones are moving towards us. The size of the points is proportional
to the magnitude of the line-of-sight peculiar velocity. "X" is  our
estimate of the direction of the tilted velocity estimated from
the SN data.} \label{mock}
\end{figure}

\section{Data Analysis}\label{data} We studied five different catalogs of
galaxy peculiar velocities to constrain the tilted velocity
$\mathbf{u}$ and the velocity dispersion $\sigma_{\ast}$. The five
data sets are (see \cite{Watkins08} for the procedure of excluding
outliers):

\begin{itemize}

\item ENEAR is a survey of Fundamental Plane (FP) \cite{FPTF}
distances to nearby early-type galaxies \cite{Costa00}. After the
exclusion of 4 outliers, there are distances to 698 field galaxies
or groups. For single galaxies, the typical distance error is $20
\%$. The characteristic depth of the sample is 29 $h^{-1}$Mpc.

\item SN are 103 Type Ia supernovae distances \cite{Tonry03},
limited to a distance of $\ltsim$150 $h^{-1}$Mpc. SN distances are typically
precise to 8\%. The characteristic depth is 32 $h^{-1}$Mpc. See
Fig.~\ref{mock}.

\item SFI++ \citep{Springob07}, based on the Tully-Fisher (TF)
\cite{FPTF} relation, is the largest and densest peculiar velocity
survey considered here. After rejection of 38 (1.4\%) field and 10
(1.3\%) group outliers, our sample consist of 2720 field galaxies
and 736 groups, so we divide it into two sub-samples, field
samples SFI++$_{\text{F}}$, and group samples SFI++$_{\text{G}}$.
The characteristic depth is 34 $h^{-1}$Mpc.

\item SMAC \cite{Hudson99} is an all-sky Fundamental Plane (FP)
\cite{FPTF} survey of 56 clusters, with characteristic depth 65
$h^{-1}$Mpc.


\item COMPOSITE is the combined catalogs (4536 data in total,
compiled in \cite{Watkins08,Feldman09}) which has the
characteristic depth of 33 $h^{-1}$Mpc. It is a combination of SN,
SFI++, SMAC, ENEAR and also the samples from SBF \cite{Tonry01}
(69 field and 23 group galaxies, characteristic depth 17 $h^{-1}$Mpc),
EFAR \cite{Colless01} (85 clusters and groups, characteristic
depth 93 $h^{-1}$Mpc), SC \cite{Giovanelli98} (TF-based survey of spiral
galaxies in 70 clusters, characteristic depth 57 $h^{-1}$Mpc), and
Willick \cite{Willick99} (Tully-Fisher based survey of 15
clusters, characteristic depth 111 $h^{-1}$Mpc).
\end{itemize}


\begin{figure*}[tbp] 
\includegraphics[bb=0 0 400 274,width=3.2in,height=2.2in]{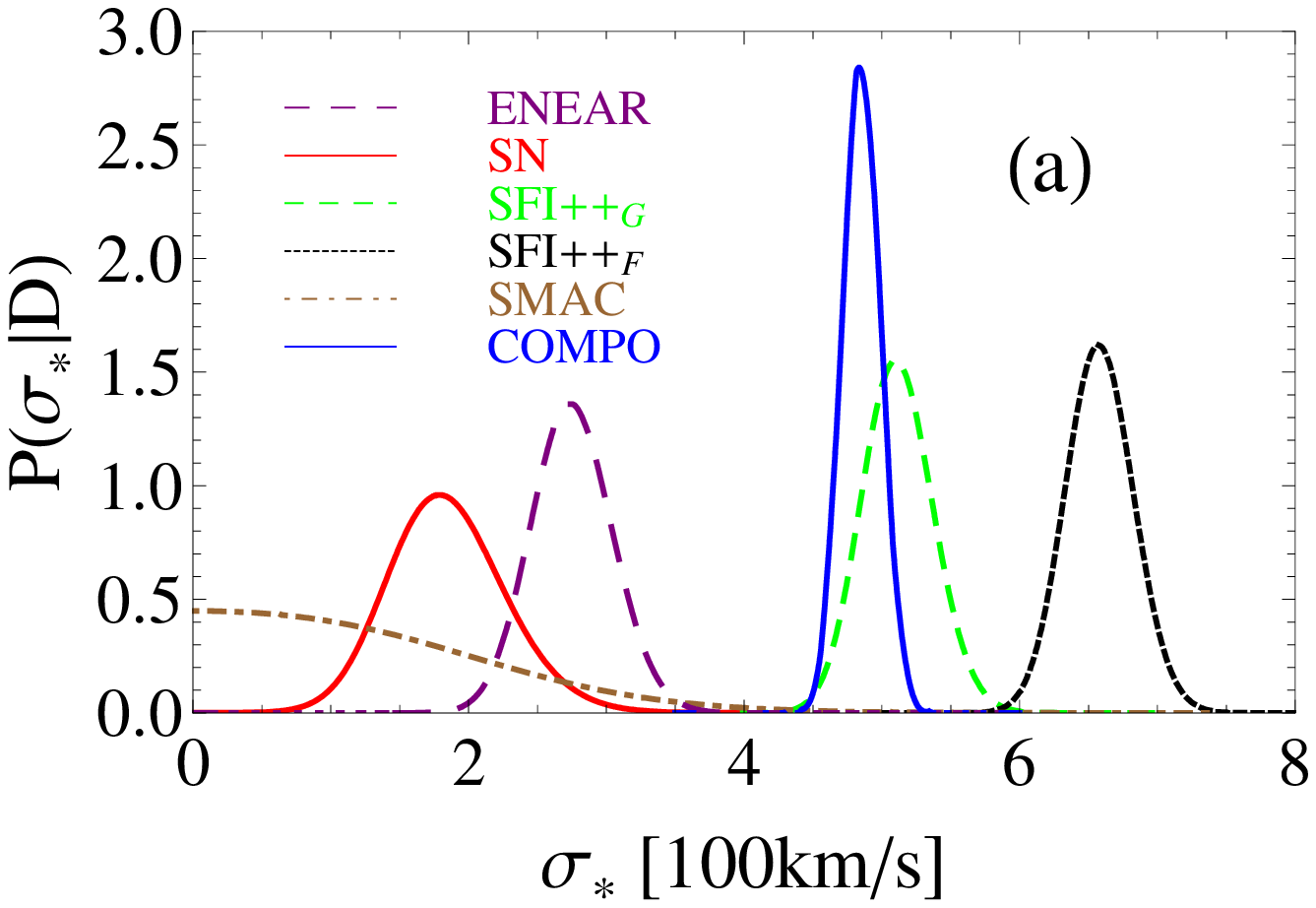} 
\includegraphics[bb=0 0 431 283,width=3.2in,height=2.2in]{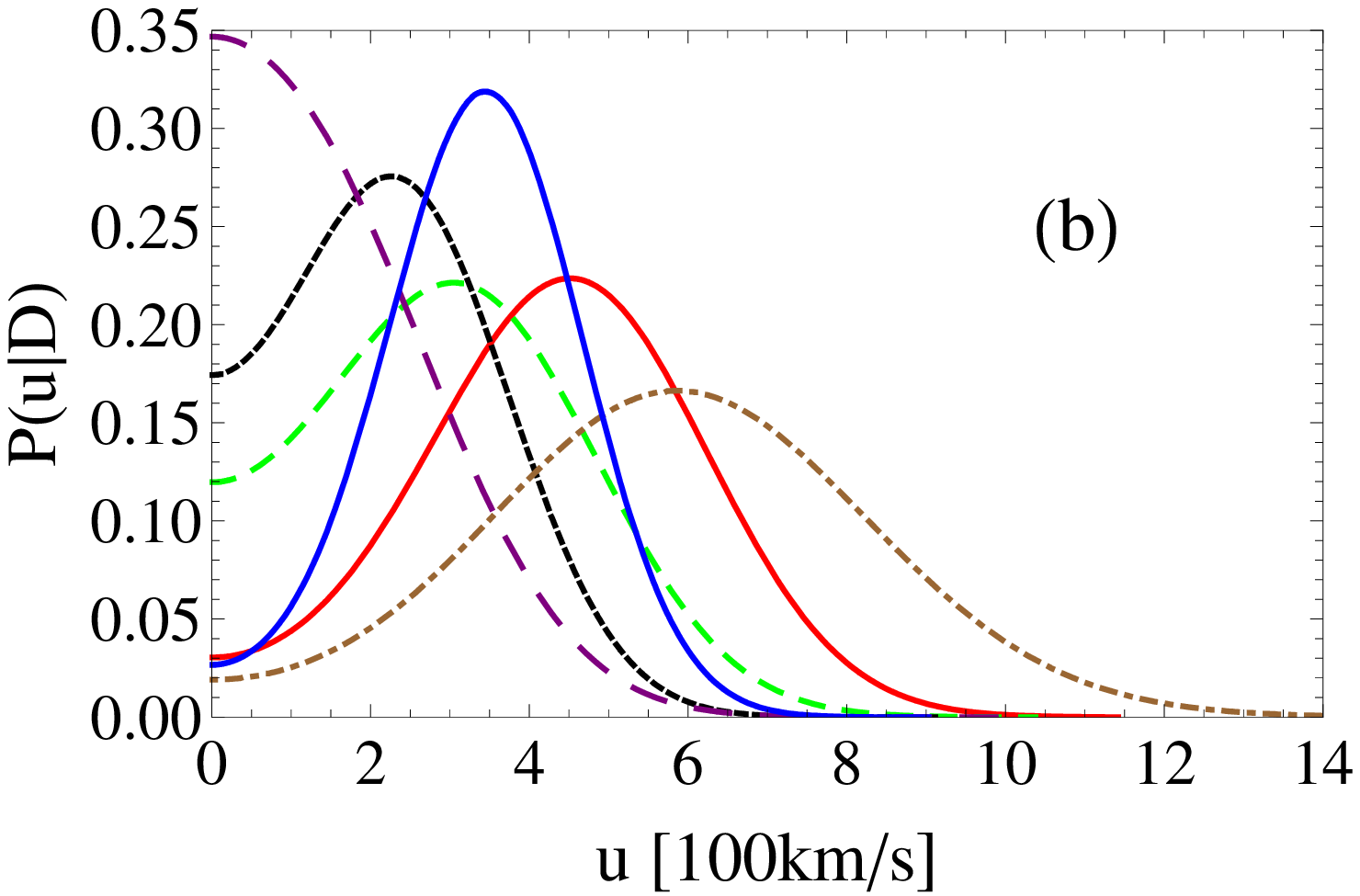} \\
\includegraphics[bb=0 0 414 280,width=3.2in,height=2.2in]{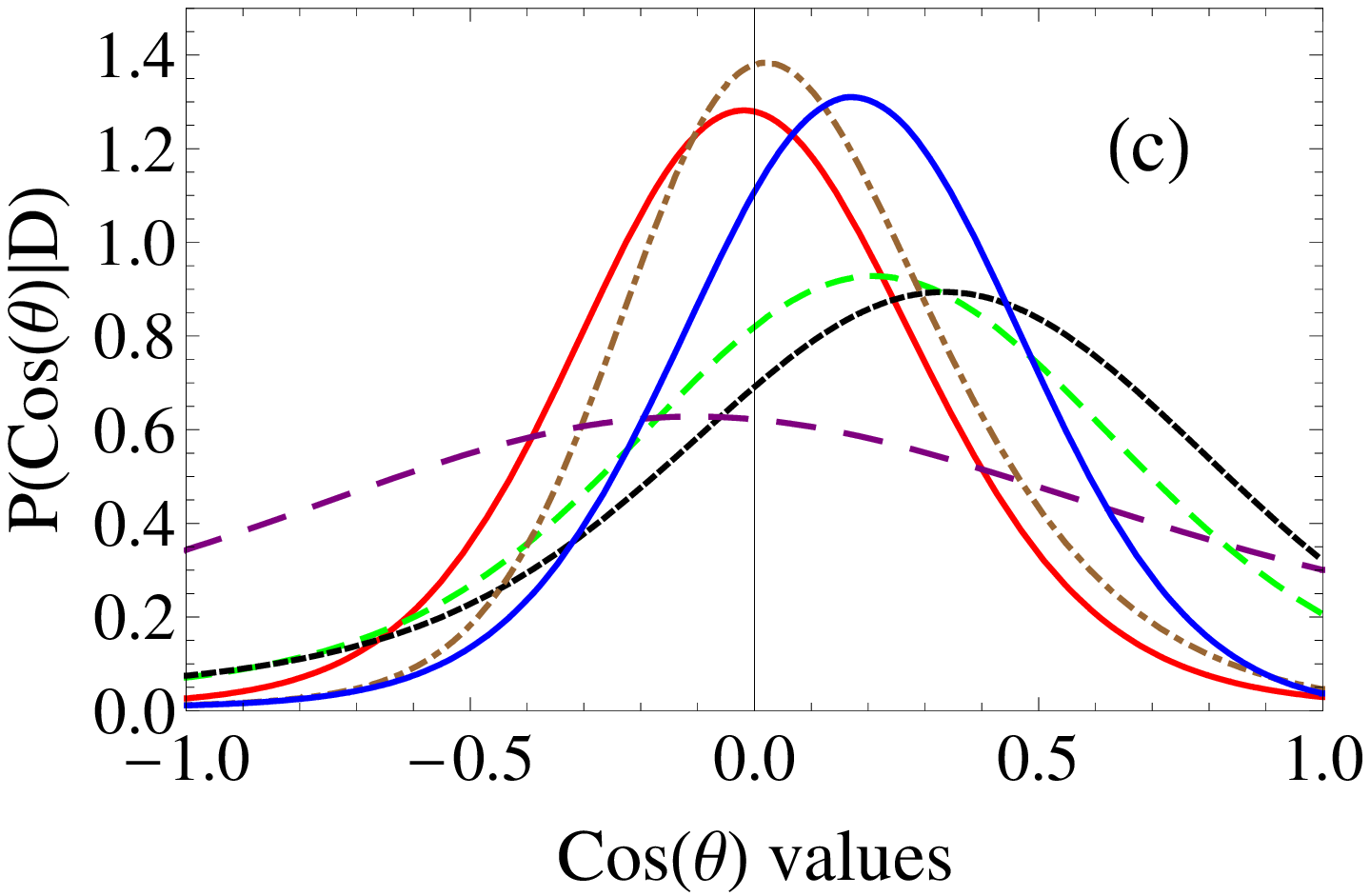} 
\includegraphics[bb=0 0 405 274,width=3.2in,height=2.2in]{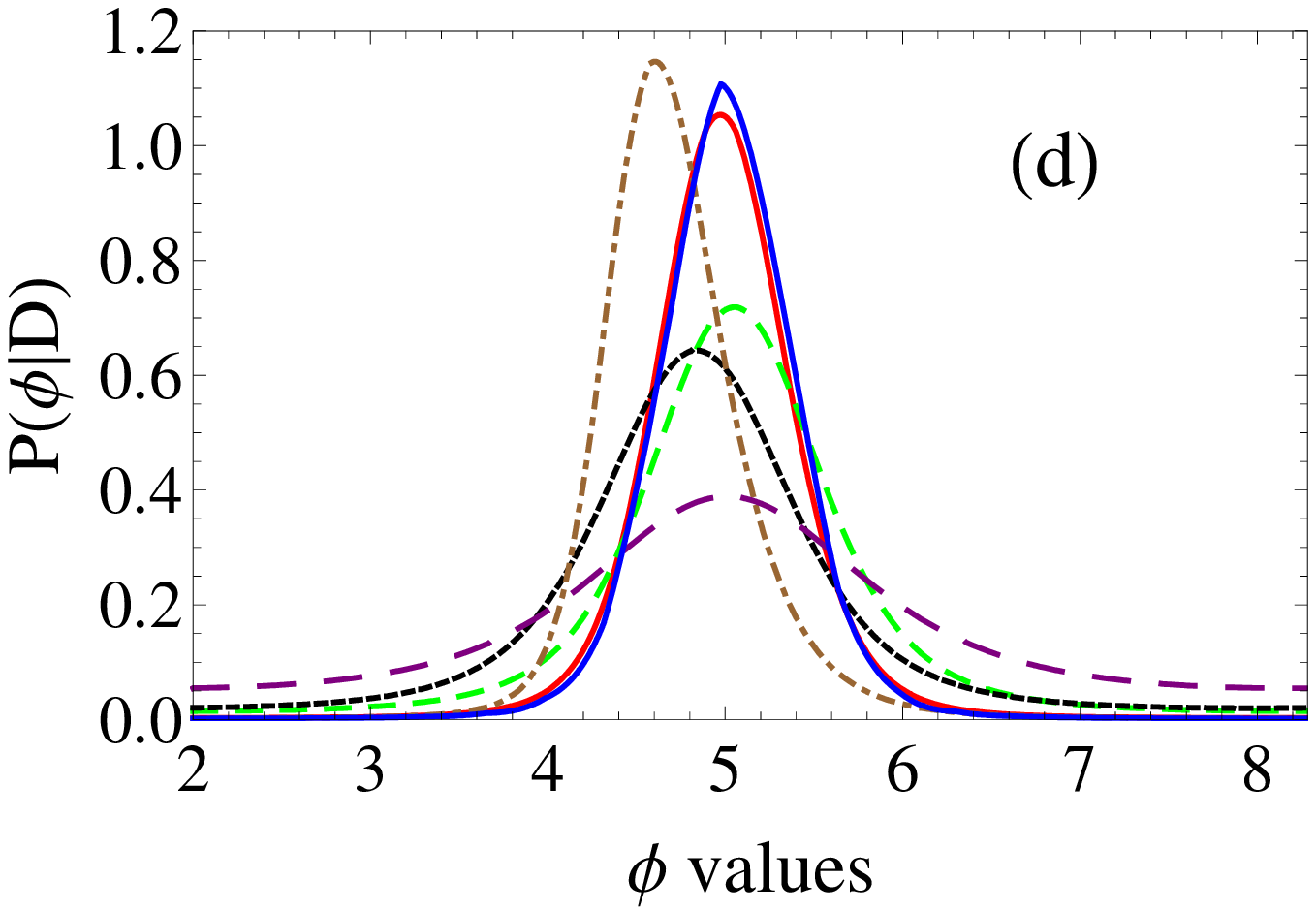}   \\
 \caption{The 1-D marginalized posterior
probability distribution functions of the parameters : (a)
$\sigma_{\ast}$, (b) magnitude of $\mathbf{u}$, (c) Cos($\theta$),
(d) $\phi$.} \label{realdata}
\end{figure*}

\begin{figure*}[tbp] 
\centerline{\includegraphics[bb=0 0 367
368,width=3.2in]{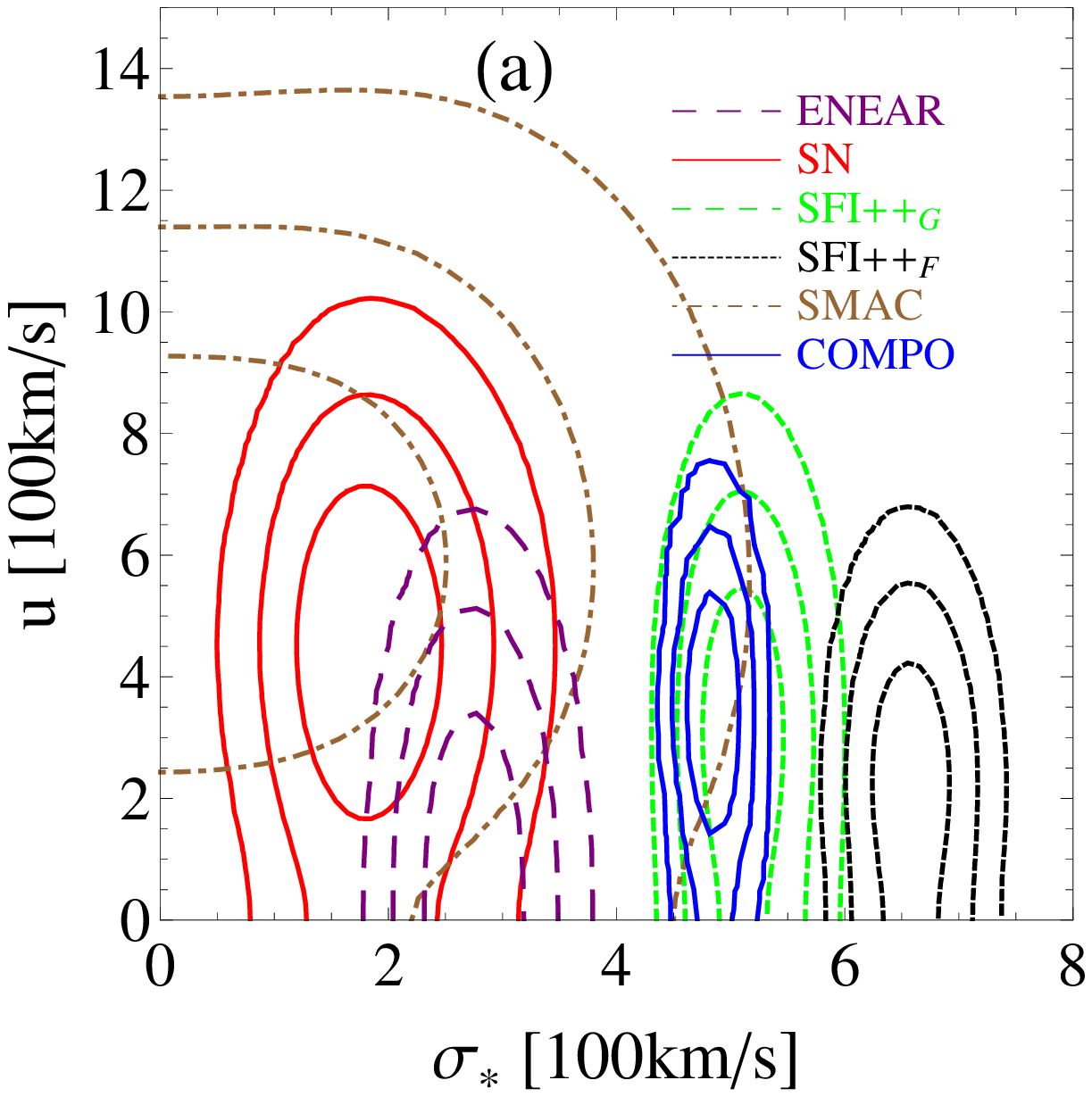}
\includegraphics[bb=0 0 362 369,width=3.2in]{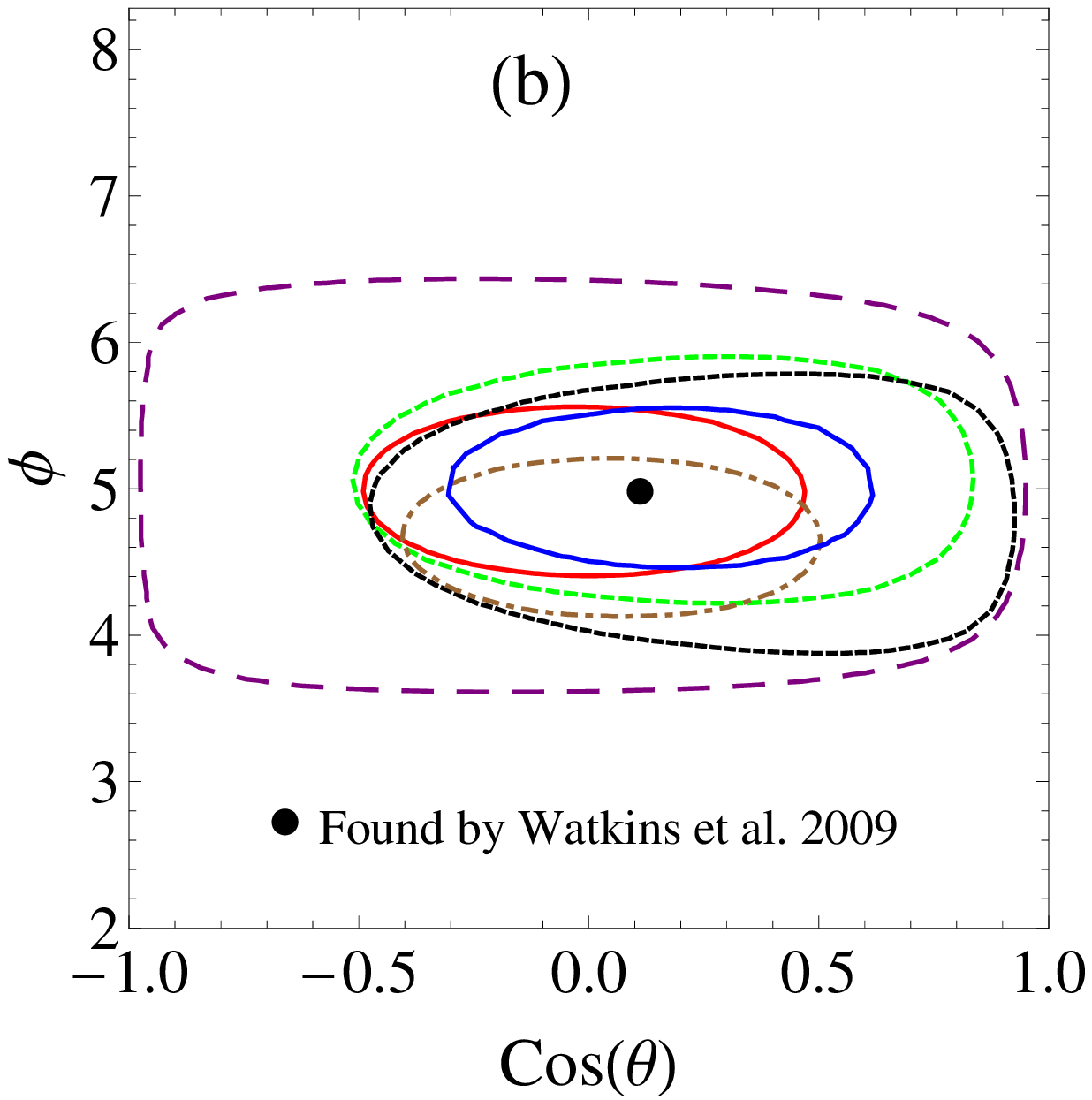}}
 \caption{(a) The 68\%, 95\% and 99.7\% Bayesian confidence interval contours  for parameters $\sigma_{\ast}$-$u$. (b) The 68\% contours for Cos($\theta$)-$\phi$.
 The black dot is the direction of
  bulk flow found by \cite{Watkins08}.}
\label{realdata2}
\end{figure*}

In Fig. \ref{realdata}, we show the marginalized 1-D posterior
probability distribution functions  of the small scale velocity
and intrinsic dispersion $\sigma_{\ast}$, and the velocity vector
$\mathbf{u}$. The best fit and marginalized error bars are listed in Table~\ref{tab1}.
In panel (a) of Fig.~\ref{realdata}, we see that different surveys
prefer different $\sigma_{*}$. For the SN catalog, there is a
smaller $\sigma_{*}$, because the supernovae light curves can be
used to calibrate the distance measurement very precisely, so the
scatter of the line-of-sight velocity is small.
 In the COMPOSITE catalog, $\sigma_{\ast}
\sim 450$ km/s reflects the average value of $\sigma_{\ast}$ of
all of the catalogs, which can be treated as the average value of
the small scale velocity  and the intrinsic dispersion. Using a
single  $\sigma_{\ast}$  should not have a significant effect, in
general, because the catalogs are of typical depth $> 40$ $h^{-1}$Mpc or
in better units $> 4000$ km/s, and the distance indicators have
typically 20\% error in the measurement which means that typically
$\sigma_n > 800$ km/s. So whether $\sigma_{\ast}$ is 400 km/s or
600 km/s is not of much consequence if $\sigma_{n}>\sigma_{\ast}$.

In  panel (b) of Fig. \ref{realdata}, we show the marginalized 1-D
distribution of the magnitude of the $\mathbf{u}$. Also, from
Table \ref{tab1} we can see that the SFI++ and ENEAR catalogs can
provide fairly tight upper bounds on $u$, and ENEAR peaks at zero
while the SFI++ catalogs contain zero velocity within $2 \sigma$.
It should be noticed that the SN and the larger COMPOSITE catalog
can provide a non-zero $2 \sigma$ lower bound on $u$, in which the
latter provides the tightest constraints. In panels (c) and (d) of
Fig. \ref{realdata}, we show the marginalized 1-D probability
distribution of the direction of the relative velocity
$(\cos(\theta),\phi)$. We see that the distribution of
$\cos(\theta)$ and $\phi$ are very close to Gaussian, and the four
catalogs  predict a very similar direction of the velocity (Fig.
\ref{realdata2} and Table \ref{tab1}).
We should also notice that the various catalogs give consistent
constraints on the tilted velocity.

In panel (a) of Fig. \ref{realdata2}, we plot 2-D contours of
$\sigma_{\ast}$ and $u$, and we find that, in the posterior
distribution, $\sigma_{\ast}$ and $u$ are not correlated.  The
reason is easy to understand: the ``tilted Universe'' velocity
$\mathbf{u}$ describes  super-horizon dipole modulation of CMB
photons, whereas $\sigma_{\ast}$ describes the sub-horizon modes
for  small scale velocities and the intrinsic  dispersion of each
individual galaxy, so they come from completely different origins
therefore without much correlation. Thus, if the instrumental
noise $\sigma_n$ has been underestimated, the distribution of
$\sigma_{\ast}$ will be shifted to smaller values, but that will
not change the constraints on $u$ significantly. Thus, our
constraint on $u$ is pretty robust with respect to the estimate of
small scale instrumental noise. In panel (b) of Fig.
\ref{realdata2}, we plot the 1$\sigma$ contour of Cos$(\theta)$
and $\phi$. We can see that the directions of tilted velocity
found by different surveys are very consistent with each other,
and also very consistent with the results from Watkins et al.
\cite{Watkins08}.

\section{Discussion} We should notice that, the direction of
the large bulk flow velocity found by \cite{Watkins08} is within
the $1\sigma$ confidence level of the tilted velocity here, which
therefore can provide a physical origin of the large bulk flow in
\cite{Watkins08}. In \cite{Feldman09} they evaluated that  the
probability of getting a higher bulk flow within $\Lambda$CDM with
WMAP7 parameter values to be less than 2\%. In our analysis, the
bulk flow of a galaxy survey should be accounted for  in the error
bars by the cosmic variance term (Eq.~\ref{vel_noi1}). So an
alternative explanation might be that there is a feature in the
matter power spectrum which would increase the cosmic variance.
However, as the shear and octupole moments of the peculiar
velocity field are not anomalously large, this is disfavored by
the data \cite{Macaulay10}.

In addition, the direction we find is consistent with that found
in \cite{Kashlinsky} which used the dipole of the kinetic
Sunyaev-Zeldovich (kSZ) measurements to estimate the bulk flow on
Gpc scales, but our magnitude is lower than what they found.
However, there is an additional level of uncertainty in converting
the kSZ dipole into a bulk flow which makes it difficult to
estimate the magnitude of the bulk flow from the kSZ dipole
\cite{Kashlinsky}. In the bulk flow approach, Ref.
\cite{Feldman09} suggests that the bulk flow comes from scales $>$
300 $h^{-1}$Mpc. If \cite{Kashlinsky} proves correct it will be strong
support for tilt explanation since the tilted velocity should be
the same regardless of the scale probed to measure it. A tilted
Universe scenario was also proposed in \cite{Kashlinsky08} to
explain the kSZ measurement.
\begin{table*}[tbp]
\begin{centering}
\begin{tabular}{|c|c|c|c|c|c|c|}\hline
 Catalogs & Characteristic Depth ($h^{-1}$Mpc) & $\sigma_{*}$ [100km/s]
 & $u$ [100km/s] & $l$ (degrees) & $b$ (degrees) & $\Delta N$ ($2\sigma$)
\\ \cline{1-7} ENEAR & $29$ & $2.8^ \pm 0.3$ & $0^{+2.2}_{\times}$ &
 $287.2^{+68.9}_{-68.3}$ & $ -3.8^{+37.5}_{-36.0}$ & $\Delta N \geq 7$
\\ \cline{1-7}  SN & $32$ & $1.8 \pm 0.4$ & $4.5^{+1.8}_{-1.9}$ &
$284.9^{+22.9}_{-22.1}$& $-1.0^{+18.8}_{-18.3}$ & $6 \leq \Delta N
\leq 9$
\\ \cline{1-7} SFI++$_{\text{G}}$ &$34$& $5.10^{+0.7}_{-0.3}$ & $3.1^{+1.6}_{-1.9}$&
$289.7 \pm 34.9$& $10.3^{+26.8}_{-25.5}$ & $\Delta N \geq 6$
\\ \cline{1-7} SFI++$_{\text{F}}$ & $34$ & $6.6^{+0.3}_{-0.2}$ & $2.3^{+1.2}_{-1.6}$ &
$276.8^{+40.1}_{-39.0}$& $15.8^{+28.4}_{-27.1}$ & $\Delta N \geq
6$
\\ \cline{1-7} SMAC &$65$ & $0.0^{+1.7}_{\times}$ &$5.9 \pm 2.4$&
$263.8^{+23.6}_{-18.5}$& $1.1^{+18.6}_{-16.0}$ & $6 \leq \Delta N
\leq 8$
\\ \cline{1-7} COMPOSITE & $33$ & $4.8^{+0.2}_{-0.1}$ & $3.4 \pm 1.3$ &
$285.1^{+23.9}_{-19.5}$& $9.1^{+18.5}_{-17.8}$ & $6 \leq \Delta N
\leq 8$
\\\hline
\end{tabular}%
\caption{The best-fit and $1\sigma$ confidence level for the
velocity dispersion $\sigma_{*}$, the magnitude ($u$) and the
direction ($l,b$) of the tilted velocity, and the $2\sigma$
constraints on the number of e-folds of inflation. The
``$\times$'' means that the value is less than zero.} \label{tab1}
\end{centering}
\end{table*}

An intrinsic dipole on the CMB sky caused by a tilted Universe
may be explained by a pre-inflationary relic isocurvature inhomogeneity. If
inflation lasts for only a few  e-folds longer than required to
solve the horizon problem, the scales that were super-horizon at
the initial point of inflation are not very far outside our
current horizon today. Inflation requires that there was initially
a fairly smooth region of order of the inflationary Hubble
horizon. A physical scale of such a region at the onset of
inflation $l = e^{p} H_{i}^{-1}$ will have the physical scale $L=
e^{P} H_{0}^{-1}$ today, where $P=p+N-N_{\rm min}$ ($N_{\rm min}$
is the minimal number of e-folds to solve the horizon problem and
is generally assumed to be around 60) \cite{Turner91}. In the
short inflationary case where $N$ is not much larger than $N_{\rm
min}$, a remnant of a pre-inflationary Universe still exists on
super-horizon scales which may be detectable. The
quadrupole effect, aka the Grishchuk-Zel'dovich effect
\cite{GriZel78}, is also part of the effect of this large scale
inhomogeneity.

An isocurvature perturbation can produce the ``tilted Universe''
effect, which arises in inflationary models due to the
perturbations of quantum fields other than the inflaton, such as
the axion, whose energy density is subdominant compared to that of
the inflaton.
There are strong constraints on sub-horizon isocurvature perturbations and thus if they
are responsible for the  tilt it would require them to be much larger on very large scales
than they are on smaller scales. Although multi-field models will not generically produce such a sharp breaking in scale invariance
or result in isocurvature modes in the later Universe, there are double inflation models which can achieve this  \cite{Langlois96}.

The dipole anisotropy is associated with the isocurvature
fluctuation as follows \cite{Turner91,Langlois96} \footnote{There
may be some factors of order unity in this equation for different models,
but they have negligibly small effect on the $\Delta
N$ constraints due to the exponential. The same holds for $\delta
\varphi/\varphi_0 \simeq 1$.}
\begin{equation}
\frac{u}{c}\simeq \frac{H_{0}^{-1}}{L} \frac{\delta
\varphi}{\varphi_{0}} , \label{vc}
\end{equation}
where $\delta \varphi$ is the fluctuation of the quantum field,
and $\varphi_{0}$ is the background field.
An inflationary scenario that results in isocurvature perturbations in the later Universe, such as in Ref. \cite{Langlois96}, is required.
Assume that at the
onset of inflation, the isocurvature quantum fluctuation satisfies
$\delta \varphi/ \varphi_{0} \simeq 1$ (Constants of
$\mathcal{O}(1)$ won't affect the results much), thus the Hubble
horizon at the onset of the inflation is $p=0$, $l=H_{i}^{-1}$,
and $L=e^{\Delta N}H_{0}^{-1}$, where
$\Delta N=N-N_{\rm min}$ is the excess number of inflationary
e-folds beyond $N_{\rm min}$. The constraints on a ``tilted
Universe'' lead to the constraints on the number of e-folds of
inflation $u/c \simeq H_{0}^{-1}/L \simeq e^{- \Delta N}$. Therefore, an observable ``tilted
Universe'' velocity requires that inflation should last a modest
number of e-folds, which should not be too long nor too short---
if inflation lasts too long, such perturbation effects would be
washed out, if inflation is too short, the dipole effect will be
too large, making the Universe over-tilted. We show the
constraints on the number of e-folds in the last column of Table
\ref{tab1}. We find that the required extra number of e-folds is
at least 6 for all catalogs, and SN, SMAC and COMPOSITE can also
provide a $2\sigma$ upper bound on $\Delta N$ given their data.

%

Note that our study  provides a similar constraint to the
Grishchuk-Zel'dovich
effect (which can arise from either adiabatic or isocurvature perturbations) which requires that the extra number
of e-folds should be greater than about 7 at the 95\% confidence
interval \cite{Castro03}.
Also, correlations in the CMB multipoles may be used
  to estimate, from the PLANCK data,  the
tilt with an error bar similar to what we obtained here
\cite{Kosowsky10}. In the future, there is the potential to use
the large number of SNe measured by the Large Synoptic Survey
Telescope (LSST)
 to constrain
$u$ with a standard deviation of about 30 km/s \cite{Gordon07}.
 So it may be
possible to explore the duration of inflation and the
pre-inflationary quantum state at quite a precise level. The
peculiar velocity field, is therefore a powerful tool to probe the
very early Universe in a manner not accessible by  CMB
observations alone.

\section{Conclusion} In this paper, we developed a model and a
statistical method to justify whether the apparent bulk flow
motion of galaxies in our surveys is due to the subtraction of the intrinsic
dipole on the CMB sky. In the conventional bulk flow scenario, the
galaxies in the local region ($\ltsim150$ $h^{-1}$Mpc) are moving
towards direction ($l=287^{o} \pm 9^{o}$, $b=8^{o} \pm 6^{o}$)
with $v=407 \pm 81$ km/s seems in tension with the $\Lambda$CDM
predictions. However, we point out that
some fraction of the CMB dipole can be intrinsic due to large
scale inhomogeneities generated by preinflationary isocurvature fluctuations, so
that in the CMB rest frame, all of the galaxies have streaming
velocity towards the particular direction, resulting in the tilted
Universe.

We modeled an intrinsic CMB dipole as a tilted velocity $\mathbf{u}$
and developed a statistical tool to constrain its magnitude and
direction. We found that (1) the magnitude of the tilted velocity
$u$ is around $400$ km/s, and its direction is close to what was
found in previous bulk flow studies. For SN, SMAC and
COMPOSITE catalogs, $u=0$ is excluded at
 about the $2.5 \sigma$ level, which can explain the
 apparent flow;
(2) there is little correlation between the tilted velocity
$\mathbf{u}$ and galaxy's small scale velocity and intrinsic
dispersion $\sigma_{\ast}$, which confirms our assumption.

Furthermore, assuming that primordial
isocurvature modes lead to an intrinsic dipole anisotropy,
constraints on the magnitude of the tilted velocity can result in
 constraints on the duration of inflation, due to the fact that
inflation can neither be too long (no dipole effect) nor too short
(very large dipole effect). Under this assumption, the constraints
on the tilted velocity require that inflation lasts at least 6
e-folds longer (at the 95\% confidence interval) than that
required to solve the horizon problem.

Finally, we should point out that if there is an intrinsic
fluctuation, the free electrons in clusters
should be able to "see" the modulation on the sky, which can be
tested through the kSZ effect. Therefore, the results from South
Pole telescope (SPT) \cite{Hall10} and Atacama cosmology telescope
(ACT) \cite{Das10,Dunkley10} may shed some light on the intrinsic
fluctuations. Such work is in progress.

\emph{Acknowledgement---} YZM thanks
George Efstathiou, Anthony Challinor and Douglas Scott
for helpful discussions.
 CG is
funded by the Beecroft Institute for Particle Astrophysics and
Cosmology. HAF was supported in part by an NSF grant
AST-0807326 and by the University of Kansas GRF and would like to thank F. Atrio-Barandela and A. Kashlinsky  for interesting comments.


\appendix
\section{Analytic formulae of correlated Window Function $f_{12}(k)$}
\label{windowfmn} The correlated window function $f_{12}(k)$
\begin{equation}
f_{12}(k)=\int \frac{d^{2}\hat{k}}{4\pi }\left( \mathbf{\hat{r}}_{1}\mathbf{%
\cdot \hat{k}}\right) \left( \mathbf{\hat{r}}_{2}\mathbf{\cdot \hat{k}}%
\right) \times \exp \left( ik\mathbf{\hat{k}\cdot }\left( \mathbf{r}_{1}-%
\mathbf{r}_{2}\right) \right),
\end{equation}
can be calculated analytically, by transforming it into harmonic
space and using the property of spherical harmonics. The final
integral should only depend on: (a) the angle $\alpha$ between
$\mathbf{r}_{1}$ and $\mathbf{r}_{2}$ (therefore $\mathbf{r}_{1}$
and $\mathbf{r}_{2}$ should be symmetric); (b) $r_{1}$ and $r_{2}$
(amplitude of the vector); (c) $k$ ($\mathbf{k}$'s amplitude).
Therefore, we can specify $r_{1}=(0,0,1),$ $r_{2}=(0,\sin \alpha
,\cos \alpha ),$ where $\alpha $ is the relative angle between
$r_{1}$ and $r_{2}.$ Therefore,
\begin{equation}
\mathbf{A}= \mathbf{r}_{1}-\mathbf{r}_{2}=(0,-r_{2}\sin \alpha
,r_{1}-r_{2}\cos \alpha ).
\end{equation}%
So its direction and amplitude become
\begin{equation}
\mathbf{\hat{A}}=\frac{1}{A}(0,-r_{2}\sin \alpha ,r_{1}-r_{2}\cos
\alpha ),
\end{equation}%
and
\begin{eqnarray}
A=[r_{1}^{2}+r_{2}^{2}-2r_{1}r_{2}\cos \alpha ]^{\frac{1}{2}}.
\end{eqnarray}%
We can set
\begin{equation}
\mathbf{\hat{k}=}(\sin \theta \cos \phi ,\sin \theta \sin \phi
,\cos \theta ),
\end{equation}%
therefore,
\begin{equation}
\mathbf{\hat{k}\cdot \hat{A}=}\frac{1}{A}\left( (-r_{2}\sin \alpha
)\sin \theta \sin \phi +(r_{1}-r_{2}\cos \alpha )\cos \theta
\right) .
\end{equation}%
Now we can use spherical harmonic function $Y_{lm}(\theta ,\phi )$
to decompose the integrand as follows.
\begin{eqnarray}
\left( \mathbf{\hat{r}}_{1}\mathbf{\cdot \hat{k}}\right) \left( \mathbf{\hat{%
r}}_{2}\mathbf{\cdot \hat{k}}\right) &=&\cos \theta (\sin \alpha
\sin \theta
\sin \phi +\cos \alpha \cos \theta )  \nonumber \\
&=&i\sqrt{\frac{2\pi }{15}}\sin \alpha \left(
Y_{2,1}+Y_{2,-1}\right) \nonumber \\
&+& \frac{%
4}{3}\sqrt{\frac{\pi }{5}}\cos \alpha Y_{2,0}+\frac{1}{3}\cos
\alpha , \label{first_integrand}
\end{eqnarray}%
\begin{equation}
\exp (ik\mathbf{\hat{k}\cdot (\mathbf{r}_{1}-\mathbf{r}_{2})}%
)=\sum_{l}i^{l}(2l+1)j_{l}(kA)P_{l}(\mathbf{\hat{k}\cdot
\hat{A}}), \label{second_integrand}
\end{equation}%
in which we can just consider $l=0,2$ two terms in the summation.
The reason is that the mixing angle between $\mathbf{k}$ and
$\mathbf{A}$ just causes the mixing between different $m$ modes in
the spherical harmonics in Eq. (\ref{second_integrand}), so the
final non $l=0$ and $l=2$ terms vanish due to the
orthogonality. Therefore, Eq. (\ref{second_integrand}) becomes
\begin{equation}
\exp (ik\mathbf{\hat{k}\cdot (\mathbf{r}_{1}-\mathbf{r}_{2})}%
)=j_{0}(kA)-5j_{2}(kA)P_{2}(\mathbf{\hat{k}\cdot \hat{A}}),
\end{equation}%
where $j_{l}(kA)$ is spherical bessel function.
\begin{eqnarray}
P_{2}(\mathbf{\hat{k}\cdot \hat{A}}) &=&\frac{1}{2}\left( 3(\mathbf{\hat{k}%
\cdot \hat{A}})^{2}-1\right)  \nonumber \\
&=&-\frac{3}{2A^{2}}\sqrt{\frac{2\pi }{15}}(r_{2}\sin \alpha
)^{2}\left(
Y_{2,2}+Y_{2,-2}\right) \nonumber \\
&+&\frac{2}{A^{2}}\sqrt{\frac{\pi }{5}}\left( (r_{1}-r_{2}\cos \alpha )^{2}-%
\frac{1}{2}(r_{2}\sin \alpha )^{2}\right) Y_{20}  \nonumber \\
&-&\frac{3}{A^{2}}(r_{2}\sin \alpha )(r_{1}-r_{2}\cos \alpha )
\nonumber \\
&\times& i\sqrt{\frac{%
2\pi }{15}}\left( Y_{2,1}+Y_{2,-1}\right) .
\end{eqnarray}%
Note that there is no zero order term in $P_{2}(\mathbf{\hat{k}\cdot \hat{A}}%
).$ Then we use the orthogonality property of $Y_{lm}$ $\int
d^{2}\mathbf{\hat{k}}Y_{lm}Y_{l^{\prime }m^{\prime }}^{\ast
}=\delta _{ll^{\prime }}\delta _{mm^{\prime }}$ and $Y_{lm}^{\ast
}=Y_{l,-m}(-1)^{m}$ and get the final result
\begin{equation}
f_{12}(k) = \frac{1}{3}\cos \alpha
(j_{0}(kA)-2j_{2}(kA)) \nonumber \\
 +  \frac{%
1}{A^{2}}j_{2}(kA)r_{1}r_{2}\sin ^{2}\alpha .
\label{integration1}
\end{equation}%
It is clear that this integration has the three properties we listed above
and the window function $f_{12}$ depends only on ($r_{1}$,$r_{2}$,$k$,$\alpha$). This is an
independent and simplified but equivalent result to Eq.(9.32) in \cite{Dodelson}.


\begin{thebibliography}{99}
\bibitem{Watkins08} R. Watkins, H. Feldman and M. Hudson, MNRAS 392, 743 (2009).

\bibitem{Feldman09} H. Feldman, R. Watkins and M. Hudson, MNRAS, 407, 2328 (2010).


\bibitem{Turner91} M. Turner, Phys. Rev. D. 44 (1991) 3737.


\bibitem{Langlois96} D. Langlois and T. Piran, Phys.Rev. D53 (1996)
2908-2919; D. Langlois, Phys.Rev.D54:2447-2450,1996.
\bibitem{Kashlinsky08}
  A.~Kashlinsky et al.,
  Astrophys.J.,{\bf 686}, L49 (2008).

\bibitem{Erdogdu06} P. Erdogdu et al., MNRAS 373 (2006) 45-64,
[astro-ph/0610005].

\bibitem{Lavaux08} G. Lavaux, R. B. Tully, R. Mohayaee, S.
Colombi, ApJ.709 (2010) 483-498, 0810.3658 [astro-ph].



\bibitem{Erickcek08}
J.~P.~Zibin and D.~Scott,
  Phys.\ Rev.\  D {\bf 78}, 123529 (2008);
  A.~L.~Erickcek, S.~M.~Carroll and M.~Kamionkowski,
  Phys.\ Rev.\  D {\bf 78}, 083012 (2008).

\bibitem[\protect\citeauthoryear{Gorski}{Gorski}{1988}]{Gorski88}
Gorski K., Astrophys.J., {\bf 332}, L7 (1988)

\bibitem{Komatsu10} E. Komatsu et. al, 1001.4538 [astro-ph].

\bibitem{Costa00} L. N. da Costa et. al, AJ, 120 (2000) 95; M. Bernardi, et. al, AJ, 123 (2002)
2990.; G. Wenger et. al, AJ, 126 (2003) 2268.

\bibitem{Tonry03} J. L. Tonry, et. al, Astrophys.J., 594 (2003) 1.

\bibitem{Springob07} C. M. Springob, et. al, ApJ Supp, 172 (2007) 599.

\bibitem{Hudson99} M. J. Hudson, PASP, 111 (1999) 57; M. J. Hudson, MNRAS, 352 (2004) 61.

\bibitem{Tonry01} J. L. Tonry et. al, Astrophys.J., 546 (2001) 681.

\bibitem{Colless01} M. Colless et. al, MNRAS 321 (2001) 277.

\bibitem{Giovanelli98} R. Giovanelli et. al, AJ, 116 (1998) 453; D. A. Dale, AJ, 118 (1999) 1489.

\bibitem{Willick99} J. A. Willick et. al, ApJ 522 (1999) 647.

\bibitem{Macaulay10}
  E.~Macaulay et al.,
  arXiv:1010.2651 [astro-ph.CO].


\bibitem{Kashlinsky}
A.~Kashlinsky, et al., Astrophys.J. {\bf 712}, L81 (2010).



\bibitem{GriZel78} Grishchuk, L.P. \& Zel'dovich, Ya.B., Astron. Zh. 55, 209 (1978).

\bibitem{Castro03}
  P.~G.~Castro, M.~Douspis and P.~G.~Ferreira,
  Phys.\ Rev.\  D {\bf 68}, 127301 (2003).

\bibitem{Kosowsky10}
  A.~Kosowsky and T.~Kahniashvili,
  arXiv:1007.4539 [astro-ph.CO];
  L.~Amendola et al.,
  arXiv:1008.1183 [astro-ph.CO].


 \bibitem{Gordon07}
  C.~Gordon, K.~Land and A.~Slosar,
  MNRAS, 387, (2008), 371.

\bibitem{Hall10} N. R. Hall, et al. ApJ. 718 (2010) 632-646, arXiv:0912.4315 [astro-ph].

\bibitem{Das10} S. Das, et al. arXiv:1009.0847 [astro-ph].

\bibitem{Dunkley10} J. Dunkley, et al. arXiv:1009.0866 [astro-ph].

\bibitem{Dodelson} S. Dodelson, \textit{Modern Cosmology},
Academic Press 2003.

\bibitem{FPTF} Fundamental Plane (FP) and Tully-Fisher (TF)
are phenomenological procedures to determine
the galaxies' intrinsic brightness and thus their distance
independently of their redshift.

\end{thebibliography}
\end{document}